# A Search for the Radio Counterpart to the March 1, 1994 Gamma Ray Burst

D. A. Frail

National Radio Astronomy Observatory, P.O. Box 0, Socorro, NM 87801

S. R. Kulkarni

California Institute of Technology, MS 105-24, Pasadena, CA 91125

K. C. Hurley

University of California, Space Sciences Lab, Berkeley, CA 94720

G. J. Fishman, C. Kouveliotou, C. A. Meegan

NASA-Marshall Space Flight Center, Huntsville, AL 35812

M. Sommer

Max-Planck Institut fur Extraterrestrische Physik, 8046, Garching, Germany

M. Boer, M. Niel

CESR, BP 4346, 31029 Toulouse Cedex, France

and

T. Cline

NASA Goddard Space Flight Center, Greenbelt, MD 20771

## ABSTRACT

We report on the results of a search for the radio counterpart to the bright $\gamma$-ray burst of March 1, 1994. Using the Dominion Radio Astrophysical Observatory Synthesis Telescope sensitive, wide-field radio images at 1.4 GHz and 0.4 GHz were made of a region around GRB 940301. A total of 15 separate radio images were obtained at each frequency, sampling a near-continuous range of post-burst timescales between 3 and 15 days, as well as 26, 47 and 99 days. We place an upper limit of 3.5 mJy on a fading/flaring radio counterpart at 1.4 GHz and 55 mJy at 0.4 GHz. Previous searches have concentrated on searching for a counterpart at only one epoch following the outburst. In contrast, the present search maintains high sensitivity over two decades of post-burst time durations. Time-variable radio emission after the initial $\gamma$-ray burst is a prediction of all fireball models, currently the most popular model for $\gamma$-ray bursts. Our observations allow us to put significant constraints on the fireball parameters for cosmological models of $\gamma$-ray bursts.

*Subject headings:* gamma rays: bursts – radio continuum: general

## 1. INTRODUCTION

Few problems in modern-day astrophysics have defied solution like the origin of $\gamma$-ray bursters (GRBs). The Burst and Transient Experiment (BATSE) experiment aboard the *Compton Gamma Ray Observatory* (CGRO) showed that the angular distribution of bursts on the sky is isotropic and that the distribution of bursts brightnesses is non-uniform (Meegan et al. 1992). The implication of these results is that we lie at the center of a bound, spherical distribution of sources whose distances are entirely unknown. Currently, the two most popular models are the extended halo model (Hartmann et al. 1994) and the cosmological model (Mao & Paczyński 1992). The lack of significant dipole and quadrapole moments in the angular distribution of the bursts severely restricts Galactic halo models (Hakkila et al. 1994). A cosmological origin for GRBs is the simplest interpretation of the data (Paczyński 1993).

While the statistics of bursts will continue to improve and new $\gamma$-ray observations (e.g. detection of emission/absorption lines, repeating bursts, lensed events, time dilation, etc) could repudiate the models, a significant breakthrough in our understanding of GRBs will likely require identifications at other wavelengths. However, in over 20 years of searching, no counterparts at other wavelengths have been unambiguously identified with a GRB. Our lack of knowledge about the nature of the GRB parent population and the emission process has resulted in a bewildering array of models and a distance scale that is uncertain by many orders of magnitude (Nemiroff 1994).

Whether GRBs are in the Galactic halo or at cosmological distances, the energy released at the site is so enormous and occurs on such a short timescale, that it is inevitable that a relativistically expanding fireball must result, regardless of the actual origin of this energy. A broad-band spectrum of emission is expected to be produced by the interaction of such a fireball with the ambient medium or by the self-shocking of its ejecta (Paczyński & Rhoads 1993, Meszaros & Rees 1993, Katz 1994). A fireball of pure photons is excluded because its predicted spectrum is thermal (Goodman 1986) whereas the observed $\gamma$-ray spectrum is highly non-thermal. However, even a small amount of baryon contamination will convert most of the fireball energy into the kinetic energy of the baryons (Shemi & Piran 1990). Gamma rays are expected to be produced when this highly relativistic blast wave is slowed down by the ambient gas (Rees & Meszaros 1992).

Many of the current theoretical papers (see above) focus on the evolution of such shocks and the details of the emission. Leaving aside the actual details we know from pure phenomenological considerations that such shocks, as with other astrophysical shocks (supernovae, novae, etc), produce broad-band emission. From very general principles, it is clear that the dominant energy from shocks evolve from high energy to radio waves as the shock evolves. Furthermore, radio emission is an effective tracer of high energy processes, with the synchrotron emission from ultra-relativistic electrons possessing energies of a few hundred Mev to tens of Gev. Paczyński & Rhoads (1993) and Katz (1994) have specifically considered the later stages of the shock evolution when radio emission is expected. Both these works indicate that detectable radio emission (a few mJy or brighter) could be seen over a period of days to weeks following the GRB event.

The work done here specifically explores this crucial parameter space and has been motivated by baryon-loaded fireball models and in general, as argued above, by models which involve shock emission. In contrast, much of the previous work has emphasized the search for *quiescent* emission associated with GRBs long after the initial burst (e.g. Hjellming & Ewald 1981, Schaefer 1992).

## 2. OBSERVATIONS

On 1 March, 1994 at 72637 s UT a burst, lasting 40 seconds was detected by the BATSE instrument. With a fluence of $3.5 \times 10^{-5}$ erg cm$^{-2}$ (50-300 keV) GRB 940301 is in the top 2% of bright bursts detected by BATSE thus far. GRB 940301 was also seen by the COMPTEL instrument on board CGRO (Kippen et al. 1994a) allowing for a better localization of the burst. An annulus of arrival for the burst was also determined from the time delay between the detection of GRB 940301 on the CGRO and Ulysses spacecraft (Hurley et al. 1994). Like another previously bright burst (Schaefer et al. 1994), GRB 940301 was the object of an extensive multi-wavelength campaign summarized by Harrison et al. (1994). Further interest in this GRB is prompted by a claim from Kippen et al. (1994b) that GRB 940301 and GRB 930704 are from overlapping locations.

The Dominion Radio Astrophysical Observatory (DRAO) Synthesis Telescope is well-suited to search for a radio counterpart to GRBs. This telescope consists of seven 9-m antennas on an east-west array with a maximum baseline of 600 meters. Three of the antennas are on a track and can be moved to provide a full sampling of the available spatial frequencies over a 12 day observing interval. Two continuum frequencies are observed simultaneously, 408 MHz (74 cm) and 1420 MHz (21 cm), producing fields of view (at 20% response) of 8.1° and 2.6°, respectively. It is notable that the field of view at 408 MHz is equal to the typical positional accuracy (error circle of 4° radius) of the BATSE detectors for bright bursts (Fishman et al. 1994).

Radio observations of a field centered on the original COMPTEL position (Kippen et al. 1994a) began at DRAO on March 4, three days following the GRB. The high declination ($\delta = +64°$) and a location well out of the Galactic plane $(l, b) = (151°, +24°)$, made GRB 940301 a favorable target for the DRAO telescope. Daily observations were made each evening, lasting approximately 12 hours, from March 4.99 UT to March 16.83 UT. Data for March 9 was not obtained due to technical difficulties. In addition to these 13 days of near-continuous monitoring, three 12-hour observations were made on March 28.05 UT, April 17.79 UT, and June 9.19 UT. In summary, these observations sampled a range of post-burst timescales between 3 and 15 days, as well as 26, 47 and 99 days.

Preliminary calibration and editing of the data was handled by the DRAO staff. Daily images were made at both frequencies, and an integrated image was made by adding all the data together.

The rms noise on an image made from a single day's data was 0.7 mJy/beam at 21 cm and 11 mJy/beam at 74 cm. The synthesized beams are approximately 1 arcmin at 21-cm and 3.3 arcmin at 74 cm.

## 3. RESULTS

In Fig. 1 we show a 21-cm image made by combining the data of all fifteen days together. There are 245 radio sources in this image with flux densities between 0.8 mJy and 110 mJy, a number that is consistent with predictions of extragalactic source counts at 1.4 GHz when weighted by the response pattern of the 9-m antennas (Windhorst, van Heerde & Katgert 1984). From previous optical/radio surveys it is known that the majority of the radio source population within this flux density range are extragalactic sources (Kron, Koo & Windhorst 1985). Spiral galaxies and giant ellipticals form 80% of the total, quasars 20%. Galactic radio stars are rare. There is no evidence for any significant local enhancement of sources in our 1.4 GHz or 0.4 GHz fields, nor do we find any cataloged clusters. Furthermore, all of the sources cataloged in the 1.4 GHz survey of Condon & Broderick (1986) of this area (above their 25 mJy confusion limit) are detected in our 21-cm image.

The search for a time-variable radio source was carried out in several different ways. The daily images at 1.4 GHz were examined for any significant flux variations over the duration of the observations. By rapidly blinking the daily images from successive days together we were able to carefully search for variable sources. This was done both with the original daily images and with images from which the mean flux density had been subtracted from each of the sources first (i.e. residual daily images). The full $3°$ degree field of view was searched by dividing the image into 9 subsections. Artificial variable sources were injected at arbitrary locations with unknown properties to test the method. In addition to the visual search a computer algorithm was employed on the images from each day to find any variable sources that were too faint to have been detected in the total image made from all 15 days. From these efforts we established that any "new" sources appearing in the daily images with a flux density variation of more than 3.5 mJy could be easily detected. This is 5-sigma above the daily rms noise of 0.7 mJy/beam. As no correction has been applied for attenuation by the primary beam this limit becomes less severe as we move outward from the pointing center (half power radius=52 arcmin). However, within the region bounded by the COMPTEL error box and the IPN arc our conclusions are still valid. Furthermore, here is no evidence for a rising or decaying radio source at this anywhere in these images.

Sources which vary substantially over an observing interval with a radio interferometer can also be recognized by the image artifact they produce (Cotton 1989). We searched for any strong intra-day variability (i.e. a bright flare within a single 12-hr observation or interstellar scattering) by looking for a circular grating sidelobe response around all sources. Apart from weak rings around the brightest sources (caused by uncertainties in the gain calibration at a level of a few percent) no such features were found.

The 245 radio sources identified in the total image received special attention. Individual light curves were extracted for all of them and they were examined in detail. No sources showed variations $\geq \pm 4\sigma$ ($\sim \pm 2.8$ mJy). Only 35 had flux density fluctuations of $\pm 3\sigma$ above the mean. While this is more than expected from pure statistical fluctuations of Gaussian noise it is consistent with the noise statistics derived from arbitrary locations in the image. With one exception the $3\sigma$ flux density fluctuations are one-time negative dips (21 cases) or positive spikes (13 cases) which occur on a single day. For all other days these source are constant. The remaining source is found within the $1\sigma$ bounds of the COMPTEL error circle but is outside the region defined by the intersection of the IPN of Hurley et al. (1994). It has a mean flux of 28 mJy. Its light curve (see Fig. 2.) shows the flux density steadily decreasing up to 10 days days after the burst (except for a single "flare" on day 6), followed by a rapid rise to a plateau. The temporal signature of this source (a decay followed by a subsequent rise on top of a threshold of about 25 mJy) is not the expected behavior of a GRB counterpart. It is likely that is a radio variable such as a BL Lac, RS CVn, etc. Gregory and Taylor (1986) concluded that 2-3% of all radio sources exhibit short-term ($\sim 21$ days) variability at 5 GHz. Further progress requires higher resolution imaging of the source and we plan to carry this out at the VLA.

The single-day variations could also occur because of a subtle instrumental effect that affects arrays with a modest number of telescopes like the DRAO array. Everyday, the array configuration was changed and thus the point spread function is different for each of our maps. Thus the integrated flux density of the sources that we recover from the images may vary. The effect is most severe for those sources with size comparable to the synthesized beam. There is some indication that this is happening. Two of the single-day variables are identified in the SIMBAD database with spiral galaxies (UGC 3511 and UGC 3577), both of which have measured angular extents of 1-2 arcmin, close to the size of the synthesized beam at 21 cm.

The limits derived from the 74-cm images are not as severe. The confusion level (i.e. the flux density contribution from the unresolved radio sources in the 3.3 arcmin beam) is comparable to instrumental rms noise level of 11 mJy/beam. There are approximately 110 sources with flux densities between 1.4 Jy and 30 mJy in the COMPTEL error circle alone. The full $9°$ field in which we searched for a variable source has close to 1000 sources. A visual search was made for variability in the entire field in the same manner as described above. Individual light curves were extracted for those sources in the inner $3° \times 3°$ region which covers the entire area of the 1.4 GHz field and encloses the COMPTEL error circle and its intersection with the IPN annulus. Variations were detected (50-75 mJy) from numerous sources well in excess of that expected from Gaussian noise fluctuations. The origin of these fluctuations is unknown but they are likely due to an instrumental effect (i.e. large day-to-day gain variations in the array). None of these sources showed any systematic brightening or fading of their flux densities. We conservatively set an upper limit of 100 mJy to the flux variations of any source at 74-cm.

## 4. DISCUSSION

A detailed examination of the 1.4 GHz images taken over 15 epochs from 3 days to 99 days after the original $\gamma$-ray burst has failed to turn up a convincing radio counterpart of GRB 940301. We place a limit of 3.5 mJy on our non-detection of a fading/flaring radio counterpart at an observing frequency of 1.4 GHz. This value is five times the radiometric noise and is clearly a stringent upper limit. Of the 245 radio sources detected in the $3° \times 3°$ field none were seen to exhibit variations of $\geq \pm 4\sigma$ above the mean but there were 35 sources seen with variations $\geq \pm 3\sigma$. This sample may contain a few real variables but the majority of the fluctuations are due to random noise or an instrumental effect. The most prudent conclusion is that in the region of interest bounded by the IPN arc and the COMPTEL error circle, any radio counterpart to GRB 940301 is below 3.5 mJy.

We interpret our null result in the framework of the Paczyński & Rhoads (1993) model. These authors predict that a GRB at a distance $d$ with a fluence $S$ will produce a short-lived radio transient whose flux density rises as $t^{5/4}$, peaking with a maximum flux density of $F_{\rm peak}$

$$F_{\rm peak} = 9\,{\rm mJy}\ C_F\,(\frac{\xi_\gamma}{0.1})^{-7/8}(\frac{d}{0.5\ {\rm Gpc}})^{-1/4}(\frac{S}{10^{-4}\ {\rm erg\ cm^{-2}}})^{7/8}(\frac{\rho}{10^{-24}\ {\rm g\ cm^{-3}}})^{1/8}(\frac{\nu}{1.4\ {\rm GHz}})^{5/8} \quad (1)$$

at a time $t_{\rm peak}$ after the initial burst

$$t_{\rm peak} = 47\,{\rm days}\ C_t\,(\frac{\xi_\gamma}{0.1})^{-1/2}(\frac{d}{0.5\ {\rm Gpc}})(\frac{S}{10^{-4}\ {\rm erg\ cm^{-2}}})^{1/2}(\frac{\rho}{10^{-24}\ {\rm g\ cm^{-3}}})^{1/2}(\frac{\nu}{1.4\ {\rm GHz}})^{-3/2} \quad (2)$$

thereafter declining as $t^{-3/4}$. The total time above half maximum is approximately $2t_{\rm peak}$. Here $C_t$ and $C_F$ are dimensionless constants that emerge from making the conservative assumption that the fireball is predominantly dominated by baryon energy density and not by electron or magnetic fields; the latter two are assumed to be of the order of 1% of the total energy density; $\xi_\gamma$ is the assumed efficiency of the fireball energy going into $\gamma$-rays; $\rho$, the ambient gas density and $\nu$, the frequency of observations.

With the normalizations adopted in Eqns (1) and (2) with $S = 3.5 \times 10^{-5}$ ergs cm$^{-2}$ for GRB 940301, we would have expected a radio transient 28 days after the $\gamma$-ray burst with $F_{\rm peak}$=3.6 mJy. No such source was seen although with our temporal sampling and flux density limits we would have been capable of detecting such time variability. These observations constrain the values of certain parameters assumed by Paczyński & Rhoads (1993) for the extragalactic model (see above). Either the assumptions about the fractional energies of the particles and field involved in the radio emission (via the parameters $\xi_\gamma$, $C_F$, and $C_t$) need to be lowered, or the density $\rho$ is smaller than assumed. Raising either of these quantities above the normalizations adopted in Eqns (1) and (2) is ruled out by our observations. While Paczyński & Rhoads (1993) made relatively conservative assumptions about the energy distribution, their value for $\rho$ may have been optimistic. We note that the hot phase of the interstellar medium, which is at least half the the volume in spiral galaxies, has $\rho \sim 10^{-27}$ g cm$^{-3}$ (Kulkarni & Heiles 1988). Detection of radio emission from a GRB in this density regime will require sensitivities at sub-mJy levels. Alternatively, as discussed by Paczyński & Rhoads (1993), the GRB energy could be highly collimated, reducing the power of the burst by several orders of magnitude and rendering the radio emission virtually undetectable.

Our 0.4 GHz observations pose only a weak constraint on the galactic halo models. Here $d \sim 10^{-4}$ Gpc and the corresponding $t_{\rm peak}$ is much smaller than a day for any reasonable value of $\rho$. Only when we assume full equipartition between the ions, electrons and magnetic fields (which make $C_F$ and $C_t$ roughly $50\times$ larger) would we expect to detect a radio transient.

GRB 940301 is the first $\gamma$-ray burst to have received a comprehensive radio monitoring effort over such a wide range of timescales. Our approach has been unique compared to past radio efforts. Rather than a single image of a small error box months or years after the initial burst, we have imaged a large field a short time after the initial burst and monitored the region over a long period of time. The resulting limits on time-variable radio emission have given us a first opportunity to test predictions of radio emission from relativistic fireball models. Given the many uncertainties in equations (1) and (2) we feel that the full range of relevant timescales and flux densities have yet to be covered, and thus a very strong case can be made for continuing radio observations of GRB 940301 and other bright GRBs despite our null result. The detection of even a single confirmed radio counterpart would be a significant advance in this field, confirming the basic fireball model and providing an accurate position for immediate follow-up at other wavelengths. Future efforts should concentrate on high radio frequencies where the flux density of the radio transient is both brighter ($F_{\rm peak} \propto \nu^{5/8}$) and reaches its maximum sooner ($t_{\rm peak} \propto \nu^{-3/2}$). Additionally, as mentioned by Paczyński & Rhoads (1993), the affect of their assuming that the bursts are spherically symmetric, is reduced by observing at higher frequencies.

## ACKNOWLEDGEMENTS

The Dominion Radio Astrophysical Observatory (DRAO) is operated as a national facility for radio astronomy by the National Research Council of Canada. DAF thanks R. S. Roger and the rest of the DRAO staff for their help with the observations, as well as A. Beasley, K. Dwarakanath and M. Rupen for useful discussions. SRK's research is supported by the NSF, NASA and the Packard Foundation. The Ulysses GRB experiment was constructed at the CESR in France with support from the Centre National d'Etudes Spatiales and at the Max-Planck Institute in Germany with support from FRG Contracts 01 ON 088 ZA/WRK 275/4-7.12 and 01 ON 88014. KH gratefully acknowledges assistance from JPL Contract 958056 and NASA Grant NAG5-1559. This research has made use of the Simbad database, operated at CDS, Strasbourg, France.

Fishman, G. J. et al. 1994, ApJS, 92, 229

Gregory, P. C & Taylor, A. R. 1986, AJ, 92, 371

Goodman, J. 1986, ApJ, 308, L47

Hakkila, J. et al. 1994, ApJ, 422, 659

Harrison, T. E. et al. 1994, (NMSU Preprint, submitted to A&A)

Hartmann, D. H. et al. 1994, ApJS, 90, 893

Hjellming, R. M. & Ewald, S. P. 1981, ApJ, 246, L137

Hurley, K. et al. 1994, IAU Circular No. 5944

Katz, J. I. 1994, ApJ, 422, 248

Kippen, R. M. et al. 1994a, IAU Circular No. 5943

Kippen, R. M. et al. 1994b, (preprint, submitted to A&A)

Kron, R. G., Koo, D. C., & Windhorst, R. A. 1985, A&A, 146, 38

Kulkarni, S. R. & Heiles, C. E. 1988 in Galactic and Extragalactic Radio Astronomy, eds. K. I. Kellerman & G. L. Verschuur, (Berlin: Springer-Verlag), 95

Mao, S. & Paczyński, B. 1992, ApJ, 388, L45

Meegan, C. A. et al. 1992, Nature, 355, 143

Meszaros, P. & Rees, M. J. 1993, ApJ, 405, 278

Nemiroff, R. J. 1994, Comments Astrophys., 17, 189

Paczyński, B. 1993 in AIP Conf. Proc. 280, Compton Gamma-Ray Observatory, eds. M. Friedlander, N. Gehrels, & D. J. Macomb (New York: AIP), p 891.

Paczyński, B. & Rhoads, J. E. 1993, ApJ, 418, L5

Rees, M. J. & Meszaros, P. 1992, MNRAS, 258, P41

Schaefer, B. E. et al. 1994, ApJ, 422, L71

Schaefer, B. E. 1992 in Gamma-Ray Bursts: Observations, Analyses and Theories, eds. C. Ho, R. I. Epstein, & E. E. Fenimore (Cambridge: Cambridge University Press), p 107

Shemi, A. & Piran, T. 1990, ApJ, 365, L55

Windhorst, R. A., van Heerde, G. M. & Katgert, P. 1985, A & A Sup., 58, 1

---

Fig. 1.— A radio image at 1.4 GHz (20 cm) of a 3° × 3° field centered on GRB 940301. The 1-$\sigma$ error radius of the COMPTEL error box is shown as well as the overlapping section of the IPN annulus.

Fig. 2.— A 1.4 GHz light curve for the only source in our sample to show $3\sigma$ variations above the mean on more than one day. No other sources were detected with evidence of a systematic brightening or fading in their light curves and no variations above $4\sigma$ were detected.